\documentclass[a4paper,12pt]{article}
\usepackage{amsmath,amssymb,rotating}
\usepackage{cite}
\newtheorem{theorem}{Theorem}
\def\pa{\partial}
 \def\bea  {\begin{eqnarray}}   \def\eea  {\end{eqnarray}}
\numberwithin{equation}{section}

\def\qed{\vrule height0.6em width0.3em depth0pt}

\title {\bf Generalized symmetry integrability test for discrete equations on the square lattice}

\author{{\bf D. Levi}
\\ {\small Dipartimento di Ingegneria Elettronica,}
\\ {\small Universit\`a degli Studi Roma Tre and Sezione INFN, Roma Tre,}
\\ {\small Via della Vasca Navale 84, 00146 Roma, Italy}
\\ {\sl\small E-mail: levi@roma3.infn.it}
\and {\bf R.I. Yamilov}
\\ {\small Ufa Institute of Mathematics, Russian Academy of Sciences,}
\\ {\small 112 Chernyshevsky Street, 450008 Ufa, Russian Federation}
\\ {\sl\small E-mail: RvlYamilov@matem.anrb.ru}}

\begin{document}
\maketitle

\begin{abstract}
We present an integrability test for discrete equations on the square lattice, which is based on the existence of a generalized symmetry. We apply this test to a number of equations obtained in different recent papers. As a result we prove the integrability of  7 equations which differ essentially from the $Q_V$ equation introduced by Viallet and thus from the Adler-Bobenko-Suris list of equations therein contained.
\end{abstract}

\section{Introduction}\label{sec1}

As it is well known, the generalized symmetry method allows one to classify integrable equations of a certain class and to test a given equation for integrability \cite{mss91,msy87}. In the case of $1+1$ partial differential equations it has been used to develope a computer PC-package DELiA, written in Turbo PASCAL by Bocharov  \cite{delia}. This program can be used  to prove integrability and compute symmetries of given evolutionary partial differential equations. The symmetry approach has also been applied with success to study the integrability of differential difference equations  \cite{asy00,y06}. Here we apply, using the  theoretical results contained in ref. \cite{ly09},  the method to partial difference equations defined on a quad--graph.
We will consider autonomous equations of the form
\begin{equation}\label{i1}
	\mathcal E (u_{n,m}, u_{n+1,m}, u_{n,m+1}, u_{n+1,m+1}) = 0 ,
\end{equation}
where $n,m$ are arbitrary integers, i.e. autonomous equations which have no explicit dependence on the point $(n,m)$ of the lattice and consequently are invariant with respect to $n$ and $m$ translations. Following ref. \cite{ly09}, we use as integrability criterion the existence of an autonomous five-point generalized symmetry:
\begin{equation}\label{i2}
	u_{n,m,t} = \mathcal G (u_{n+1,m}, u_{n-1,m}, u_{n,m}, u_{n,m+1}, u_{n,m-1}) ,
\end{equation}
where $t$ denotes the group parameter. Due to the  complexity of the discrete case, we limit ourselves to  the simplest nontrivial case given by symmetries of the form (\ref{i2}). Moreover we assume that the symmetry is autonomous as so is the equation. As we shall see autonomous symmetries of the form (\ref{i2}) are  general enough to cover a wide class of integrable equations. Theoretically, however, an integrable equation may have a  symmetry depending on more lattice points and having lattice dependent coefficients.

We assume that nontrivial nonlinear partial difference equations which have generalized symmetries of the form (\ref{i2}) are integrable. However there might be classes of nonlinear equations contained in (\ref{i1}) which are integrable, but have no symmetry of the form (\ref{i2}). We define as {\sl trivial} a nonlinear lattice equation  (\ref{i1}) which is  factorizable or reducible by summation to a simpler equation, maybe non--autonomous, which depends on a lower number of lattice points. Example of trivial equations are
\bea \nonumber
\mathcal E &=& \omega(u_{n,m}, u_{n,m+1}) \omega'(u_{n+1,m}, u_{n+1,m+1})=0, \\ \nonumber  \mathcal E &=& \omega(u_{n+1,m}, u_{n+1,m+1}) - \omega(u_{n,m}, u_{n,m+1})=0,
\eea
with $\omega$ and $\omega'$  given functions of their arguments.

Symmetry related integrability tests can be obtained also considering approaches different from the one presented in this paper.
\begin{enumerate}
\item One can construct integrability tests based on symmetries which are given by  lattice-dependent equations, not necessarily related with master-symmetries. An example of such a symmetry is given by lattice-dependent Volterra type equations studied in \cite{ly97}. Lattice-dependent discrete equations possessing such symmetries have been recently presented in ref. \cite{xp09}.

\item An integrability test can also be obtained considering B\"acklund transformations instead of generalized symmetries. These have been used, for example, to study equations of KdV/sine-Gordon type (see \cite{abs03}, for instance).
\end{enumerate}
Alternative methods for testing and classifying difference equations not strictly related to symmetries are:
\begin{enumerate}

\item The 3D-consistency method firstly introduced as a kind of Bianchi identity related to Bianchi commutativity theorem \cite{ns97}. As a property of maps it was firstly proposed in an article by Nijhoff, Ramani, Grammaticos and Ohta \cite{nrgo01}. It has been used with success by Adler Bobenko and Suris to classify some classes of equations on the quad-graph \cite{abs03,abs09}.

\item
Grammaticos, Ramani, and Papageorgiou, \cite{grp91} proposed in 1991 the Singularity Confinement Criterion. Later on it was shown that the singularity confinement was not sufficient to prove integrability and Viallet and Hietarinta \cite{vh00} introduced the  algebraic entropy, slow growth of complexity, as a way to test  both S- and C-integrable equations.

\item Most recently the study of the growth of the so-called characteristic algebra has been applied to the case  of difference equations and have provided a way to identify and classify integrable equations on the lattice \cite{hg10b,hg10a}

\item The existence of integrals of partial difference equations, in  both space and time directions, provide a way to test and classify  linearizable difference and partial-difference equations \cite{as99,h05}. The equations which satisfy this test have  been called {\it Darboux integrable equations}.
\end{enumerate}
A recent review of techniques used to show integrability of  difference equations can be found in \cite{h11}.

In Section \ref{sec2} we review the necessary theoretical results contained in ref.  \cite{ly09}. Then we explain how, using those results, we can test any given equation (\ref{i1}) for integrability and we apply it to a well known integrable example. In Section \ref{sec3} we apply this testing tool to a number of equations which can be found in the recent literature on this subject \cite{abs09,hiv07,hyv10,lEA,ly09U}. In Section \ref{sec4} we collect together and discuss all the equations of the form (\ref{i1})  which we have shown  to satisfy the generalized symmetry test. In particular we write down 8 integrable equations which are not contained in $Q_V$ \cite{v09} and thus are not related to the  equations of the Adler-Bobenko-Suris list \cite{abs03}.

\section{Theoretical results}\label{sec2}

As eq. (\ref{i1}) and its symmetry (\ref{i2}) are autonomous, in all generality, applying the translation invariance, we can consider them at the point $n=m=0$:
\begin{equation}\label{a1}
	\mathcal E (u_{0,0}, u_{1,0}, u_{0,1}, u_{1,1}) = 0 ,
\end{equation}
\begin{equation}\label{a2}
	u_{0,0,t} = g_{0,0} = \mathcal G (u_{1,0}, u_{-1,0}, u_{0,0}, u_{0,1}, u_{0,-1}) .
\end{equation}
Eq. (\ref{a1}) must satisfy the following conditions:
\begin{equation}\label{a3}
	(\mathcal E_{u_{0,0}} , \mathcal E_{u_{1,0}} , \mathcal E_{u_{0,1}} , \mathcal E_{u_{1,1}} )\ne 0 ,
\end{equation}
where indexes  in (\ref{a3}) denote partial derivatives. These conditions are not sufficient to rule out trivial equations. The equation:
\[
(u_{0,0} + u_{1,0}) (u_{0,1} + u_{1,1} + 1) = 0
\]
provide an example of equation which is degenerate but satisfies (\ref{a3}).

We require that eq. (\ref{a1})  be  rewritable  in the form
\begin{equation}\label{a4}
	u_{1,1} = f ^{(1,1)}(u_{1,0}, u_{0,0}, u_{0,1}) ,
\end{equation}
where \bea \label{d1}
(f_{u_{1,0}}^{(1,1)}, f_{u_{0,0}}^{(1,1)}, f_{u_{0,1}}^{(1,1)}) \ne 0 ,
\eea
 and the apex $^{(1,1)}$ indicate that the function $f$ is obtained from (\ref{a1}) by explicitating  the function $u$ in the point $(n+1,m+1)$. The conditions (\ref{d1}) are necessary conditions to prevent triviality of the equation (\ref{a1}).

It is well known \cite{hr,lpsy08,ttx07} that all the equations  (\ref{a1}) classified by Adler Bobenko and Suris  have  two symmetries of the form
\begin{equation}\label{a5}
	u_{0,0,t_1} = \Phi (u_{1,0}, u_{0,0}, u_{-1,0}) , \qquad u_{0,0,t_2} = \Psi (u_{0,1}, u_{0,0}, u_{0,-1}) ,
\end{equation}
where $(\Phi_{u_{1,0}}, \Phi_{u_{-1,0}}, \Psi_{u_{0,1}}, \Psi_{u_{0,-1}}) \ne 0$. Obviously, if (\ref{a5}) is valid,  we can construct a symmetry $u_{0,0,t} = \Phi + \Psi$  of the form (\ref{a2}) with
\begin{equation}\label{a6}
(\mathcal G_{u_{1,0}}, \mathcal G_{u_{-1,0}}, \mathcal G_{u_{0,1}}, \mathcal G_{u_{0,-1}}) \ne 0 .
\end{equation}
One can also prove in all generality for a symmetry of the form (\ref{a2})  (see details and references in \cite{ly09}) that its right hand side must be of the form
\[
\mathcal G = \Phi (u_{1,0}, u_{0,0}, u_{-1,0}) + \Psi (u_{0,1}, u_{0,0}, u_{0,-1}) .
\]

So to get a coherent result we need to add to the  generalized symmetry equation (\ref{a2}) the conditions (\ref{a6}), from now on called {\sl non-degeneracy conditions}.

In \cite{x09} the author considered affine linear equations of form (\ref{a1}), (\ref{a3}), where the function $\mathcal E$ possesses the Klein  symmetry:
\begin{equation}\label{b12}
\begin{array}{l}
 \mathcal E (u_{0,0}, u_{1,0}, u_{0,1}, u_{1,1}) = \pm \mathcal E (u_{1,0}, u_{0,0}, u_{1,1}, u_{0,1}) , \\
 \mathcal E (u_{0,0}, u_{1,0}, u_{0,1}, u_{1,1}) = \pm \mathcal E (u_{0,1}, u_{1,1}, u_{0,0}, u_{1,0}) .
\end{array}
\end{equation}
We will call  equations possessing the symmetry (\ref{b12}) {\sl Klein type equations}. It has been proved in \cite{x09} that Klein type equations possess 2 nontrivial generalized symmetries of form (\ref{a5}), i.e. satisfy our test.

The class of Klein type equations contains the $Q_V$ equation \cite{v09}, see Appendix A, which is equivalent up to $Q_4$, up to M\"obius  transformations  \cite{abs03}, see e.g. \cite{x09}. The definition of the class of Kein type equations is constructive and easy to check. Moreover Klein type equations are invariant under M\"obius (linear-fractional) transformations:
\begin{equation}\label{b13}
u_{n,m} = \frac{\alpha \hat u_{n,m} + \beta}{\gamma \hat u_{n,m} + \delta} ,
\end{equation}
and it is easy to check if an equation cannot be M\"obius transformed  into the $Q_V$ equation or into an equation of the ABS list.

Eq. (\ref{a2}) is a generalized symmetry of eq. (\ref{a4}) if the following compatibility condition is satisfied:
\[
\frac{d (u_{1,1} - f^{(1,1)})}{dt} \biggl|_{[u_{1,1} = f^{(1,1)}, \; u_{0,0,t}=g_{0,0}]}= 0 .
\]
 Explicitly the compatibility condition reads:
\begin{equation}\label{a7}
[g_{1,1} - ( g_{1,0} \pa_{u_{1,0}} + g_{0,0} \pa_{u_{0,0}} + g_{0,1} \pa_{u_{0,1}} ) f^{(1,1)}] \biggl|_{u_{1,1} = f^{(1,1)}}=0,
\end{equation}
where $g_{n,m} = T_1^n T_2^m g_{0,0}$, and $T_1,T_2$ are the shift operators acting on the first and second indexes, respectively, i.e. $T_1 g_{0,0}=g_{1,0}$ and $T_2 g_{0,0}=g_{0,1}$.

To be able to check the compatibility condition between (\ref{a1}) and  (\ref{a2})  we need to define the set of independent variables in terms of which (\ref{a7}) can be split in an overdetermined system of independent equations.   In this work we choose the functions
\begin{equation}\label{a8}
 u_{n,0} , u_{0,m}
\end{equation}
as {\it independent variables}. Then, using  (\ref{a4}), all the other functions $u_{n,m}$ can be explicitly written in terms of the independent variables (\ref{a8}). So we require that eq. (\ref{a7}) is satisfied identically for all values of independent variables. Taking into account the form of $\mathcal G$, eq. (\ref{a7})  depends on the variables $u_{1,1}, u_{-1,1}, u_{1,-1}, u_{2,1}, u_{1,2}$, and this is the reason  why  (\ref{a7}) turns out to be a rather complicated functional-difference equation for the functions $f^{(1,1)} $ and $g_{0,0}$ appearing in   (\ref{a2}, \ref{a4}).

In \cite{ly09} we have proven the following theorem:

\begin{theorem}\label{te}
If (\ref{a4}) possesses a generalized symmetry of the form (\ref{a2}),  then its solutions  must satisfy the following  conservation laws
\begin{equation}\label{a9}
 (T_1 -1) p_{0,0}^{(k)} = (T_2 -1) q_{0,0}^{(k)} ,
\end{equation}
where
\par
if $\mathcal G_{u_{1,0}} \ne 0$, then
\begin{equation}\label{a10}
 p^{(1)}_{0,0} = \log f_{u_{1,0}}^{(1,1)} , \qquad
 q^{(1)}_{0,0} = Q^{(1)} (u_{2,0},u_{1,0},u_{0,0}) ;
\end{equation}
\par
if $\mathcal G_{u_{-1,0}} \ne 0$, then
\begin{equation}\label{a11}
 p^{(2)}_{0,0} = \log \frac{f^{(1,1)}_{u_{0,0}}}{f^{(1,1)}_{u_{0,1}}} , \qquad
 q^{(2)}_{0,0} = Q^{(2)} (u_{2,0},u_{1,0},u_{0,0}) ;
\end{equation}
\par
if $\mathcal G_{u_{0,1}} \ne 0$, then
\begin{equation}\label{a12}
 q^{(3)}_{0,0} = \log f^{(1,1)}_{u_{0,1}} , \qquad
 p^{(3)}_{0,0} = P^{(3)} (u_{0,2},u_{0,1},u_{0,0}) ;
\end{equation}
\par
if $\mathcal G_{u_{0,-1}} \ne 0$, then
\begin{equation}\label{a13}
 q^{(4)}_{0,0} = \log \frac{f^{(1,1)}_{u_{0,0}}}{f^{(1,1)}_{u_{1,0}}} , \qquad
 p^{(4)}_{0,0} = P^{(4)} (u_{0,2},u_{0,1},u_{0,0}) .
\end{equation}
\end{theorem}

Let us analyze in detail  Theorem \ref{te} when $k=1$.  In this case we just require $\mathcal G_{u_{1,0}} \ne 0$, while the dependence of $\mathcal G$ on the other variables is not important. In this case Theorem \ref{te} tells us that  there exists a conservation law with a function
$p^{(1)}_{0,0}$   completely defined by  (\ref{a4}) while  $q^{(1)}_{0,0}$ is an arbitrary function of  $u_{2,0}, u_{1,0}, u_{0,0}$. The other three cases are similar.

Therefore Theorem \ref{te} provides four {\it integrability conditions} in the form of conservation laws. In the case of a non-degenerate symmetry (\ref{a2}, \ref{a6}), all these integrability conditions must be satisfied.

Let us notice that, from the existence of generalized symmetries, we can easily derive many integrability conditions of this kind. Such conditions have been written down in \cite{mwx10}. However, the other conditions are in general more complicated, and more difficult to use in practice.
Work is in progress to check if these further integrability conditions, obtained requiring the existence of a recursion operator for the symmetries, allow us to get new integrable equations of this class or they just correspond to the existence of higher symmetries.

If the integrability conditions given in Theorem \ref{te} are satisfied, then there must exist functions  $q^{(1)}_{0,0}$, $q^{(2)}_{0,0}$, $p^{(3)}_{0,0}$, $p^{(4)}_{0,0}$ local in their argument. Once we know all  such functions, we can check if some autonomous five-point symmetries might exist. In fact in such a case we can construct  the four partial derivatives of $\mathcal G$ \cite{ly09}
\bea\label{a14}
 \mathcal G_{u_{1,0}} &= \exp(-q^{(1)}_{-1,0}) , \qquad \mathcal G_{u_{-1,0}} & = \exp(q^{(2)}_{-1,0}) ,  \\ \label{a14b}
 \mathcal G_{u_{0,1}} &= \exp(-p^{(3)}_{0,-1}) , \qquad \mathcal G_{u_{0,-1}} &= \exp(p^{(4)}_{0,-1}) .
\eea
These partial derivatives must be compatible
\begin{equation}\label{a14*}
 \mathcal G_{u_{1,0},u_{-1,0}} = \mathcal G_{u_{-1,0},u_{1,0}} , \qquad
 \mathcal G_{u_{0,1},u_{0,-1}} = \mathcal G_{u_{0,-1},u_{0,1}} .
\end{equation}
Eq. (\ref{a14*}) is an additional integrability condition. If this further integrability condition is satisfied we can construct $\mathcal G$ in the form
\begin{equation}\label{a15}
\mathcal G = \Phi (u_{1,0}, u_{0,0}, u_{-1,0}) + \Psi (u_{0,1}, u_{0,0}, u_{0,-1}) + \nu(u_{0,0}) ,
\end{equation}
where $\Phi$ and $\Psi$ are  known functions of their arguments while $\nu$ is an unknown arbitrary function which may correspond to a Lie point symmetry of the equation. The function $\nu$ can be  specified by considering  the compatibility condition (\ref{a7}), the last and most fundamental integrability condition.

The first  problem is to check the integrability conditions given in Theorem \ref{te}. In the case of differential difference equations  we had a similar situation, i.e. the integrability conditions were given by conservation laws depending on arbitrary functions of a limited number of variables. However  such problem was easier to solve as all discrete variables were independent and we could use the variational derivative to check them  \cite{ly09}.
Here the calculation of the variational derivative is not sufficient to prove if a given expression is a conservation law.

So, in the following, we present  a scheme for solving this problem for any given equation of form (\ref{a4}), i.e. we show how we can solve eqs. (\ref{a9})--(\ref{a13}) to obtain $q^{(1)}_{0,0}$, $q^{(2)}_{0,0}$, $p^{(3)}_{0,0}$, $p^{(4)}_{0,0}$ as  local functions of their argument. We will split the explanation into two steps.

\paragraph{Step 1.}
At first we consider the integrability conditions (\ref{a9}) corresponding to $k=1,2$.
The unknown functions in the right hand side of (\ref{a9}) when $k=1,2$ contain the dependent variable $u_{2,1}$ which, from (\ref{a4}), depends on $u_{2,0},\, u_{1,0}, \, u_{1,1}$ and thus it is not immediately expressed in terms of independent variables, but give rise to extremely complicated functional expressions of the independent variables. We can avoid this problem by applying the operators $T_1^{-1}$, $T_2^{-1}$ to  (\ref{a9}). In this case we have:
\begin{equation}\label{a16}
\begin{array}{l}
 p^{(k)}_{0,0} - p^{(k)}_{-1,0} = q^{(k)}_{-1,1} - q^{(k)}_{-1,0} = \\
 Q^{(k)}(u_{1,1},u_{0,1},u_{-1,1}) - Q^{(k)}(u_{1,0},u_{0,0},u_{-1,0}) ,
\end{array}
\end{equation}
\begin{equation}\label{a17}
\begin{array}{l}
 p^{(k)}_{0,-1} - p^{(k)}_{-1,-1} = q^{(k)}_{-1,0} - q^{(k)}_{-1,-1} = \\
 Q^{(k)}(u_{1,0},u_{0,0},u_{-1,0}) - Q^{(k)}(u_{1,-1},u_{0,-1},u_{-1,-1}) .
\end{array}
\end{equation}
Here $p^{(k)}_{i,j}$ are known functions expressed in term of (\ref{a4}). The functions $q^{(k)}_{i,j}$ are unknown, and ($q^{(k)}_{-1,1}$, $q^{(k)}_{-1,-1}$) contain the dependent variables $u_{1,1}$, $ u_{-1,1}$, $u_{1,-1}$, $ u_{-1,-1}$. Our aim is to derive from (\ref{a16}, \ref{a17}) a set of equations for the unknown function, $q^{(k)}_{-1,0}$.

To do so let us extract from  (\ref{a1}) three further expressions of the form of (\ref{a4}) for the dependent variables contained in  (\ref{a16}, \ref{a17}):
\bea\label{a18}
 && \qquad \qquad  \qquad u_{-1,1} = f^{(-1,1)}(u_{-1,0},u_{0,0},u_{0,1}) , \\ \nonumber
 &&u_{1,-1} = f^{(1,-1)}(u_{1,0},u_{0,0},u_{0,-1}) , \quad
 u_{-1,-1} = f^{(-1,-1)}(u_{-1,0},u_{0,0},u_{0,-1}) .
\eea
All functions $f^{(i,j)}$ have a nontrivial dependence on all their variables, as it is the case of $f^{(1,1)}$, and are expressed in terms of independent variables. Let us introduce the two differential operators:
\begin{equation}\label{a19}
A = \pa_{u_{0,0}} - \frac{f^{(1,1)}_{u_{0,0}}}{f^{(1,1)}_{u_{1,0}}} \pa_{u_{1,0}} -
\frac{f^{(-1,1)}_{u_{0,0}}}{f^{(-1,1)}_{u_{-1,0}}} \pa_{u_{-1,0}} ,
\end{equation}
\begin{equation}\label{a20}
B = \pa_{u_{0,0}} - \frac{f^{(1,-1)}_{u_{0,0}}}{f^{(1,-1)}_{u_{1,0}}} \pa_{u_{1,0}} -
\frac{f^{(-1,-1)}_{u_{0,0}}}{f^{(-1,-1)}_{u_{-1,0}}} \pa_{u_{-1,0}} ,
\end{equation}
 chosen in such a way to annihilate the functions $q^{(k)}_{-1,1}$ and $q^{(k)}_{-1,-1}$,  namely, $A q^{(k)}_{-1,1} = 0$, $B q^{(k)}_{-1,-1} = 0$. Applying $A$ to (\ref{a16}) and $B$ to (\ref{a17}), we obtain two equations for the unknown $q^{(k)}_{-1,0}$:
\begin{equation}\label{a21}
A q^{(k)}_{-1,0} = r^{(k,1)} , \quad B q^{(k)}_{-1,0} = r^{(k,2)} ,
\end{equation}
where $r^{(k,1)}$, $r^{(k,2)}$ are some explicitly known functions of (\ref{a4}). Considering  the standard commutator of $A$ and $B$, $ [A,B] = AB - BA$, we can add a further  equation
\bea \label {a21a}
 [A,B] q^{(k)}_{-1,0} = r^{(k,3)}.
 \eea

 Eqs. (\ref{a21}, \ref{a21a}) represent a linear partial differential system of three equations for the unknown $q^{(k)}_{-1,0}=Q^{(k)}(u_{1,0}, u_{0,0}, u_{-1,0})$. For the three partial derivatives of $q^{(k)}_{-1,0}$, this is  just a linear algebraic system of three equations in three unknown. In most of the examples considered below, this system is non-degenerate and thus it provides one and only one solution for the three derivatives of $q^{(k)}_{-1,0}$.
In these cases we can find  in unique way the partial derivatives of $q^{(k)}_{0,0}$. Then we can check the consistency of the partial derivatives and, if satisfied, find $q^{(k)}_{0,0}$ up to an arbitrary constant. Finally we check the integrability condition (\ref{a9}) with $k=1,2$ in any of the equivalent forms (\ref{a16}) or (\ref{a17}).

The non-degeneracy of the system (\ref{a21}, \ref{a21a}) depends on  (\ref{a4}) only.
So, if we have checked the non-degeneracy for  $k=1$, we know that
this is also true for  $k=2$ and vice versa. So both functions $q^{(1)}_{0,0}$, $q^{(2)}_{0,0}$ are found in unique way up to a constant of integration.

If the system (\ref{a21}, \ref{a21a}) is degenerate, the functions $q^{(1)}_{0,0}$, $q^{(2)}_{0,0}$ are defined up to some arbitrary functions. In this case the checking of the integrability conditions (\ref{a9}) may be more difficult.

In principle the  coefficients of the system (\ref{a21}, \ref{a21a}) may depend, in addition to the natural variables $u_{0,0}$, $u_{1,0}$, $u_{-1,0}$, entering in $q^{(k)}_{-1,0}$, on the independent variables $u_{0,1}$, $u_{0,-1}$. In such a case we have to require that the solution $q^{(k)}_{-1,0}$ does not depend on them. In this case  we have to split the equations of the system (\ref{a21}, \ref{a21a}) with respect to the various powers of the independent  variables $u_{0,1}$, $u_{0,-1}$, if (\ref{a4}) is rational,
 and obtain an overdetermined system of  equations for $q^{(k)}_{-1,0}$. Moreover, overdetermined systems of equations are usually simpler to solve.  There will be some examples of this kind in Section \ref{sec3}. \qed

\paragraph{Step 2.}
Let us consider now the conditions (\ref{a9}) with $k=3,4$. In this case we have a  similar situation. By appropriate shifts we rewrite (\ref{a9}) in the two equivalent forms:
\bea\label{a22}
&& p^{(k)}_{1,-1} - p^{(k)}_{0,-1} = q^{(k)}_{0,0} - q^{(k)}_{0,-1} = \\ \nonumber
 && \quad P^{(k)}(u_{1,1},u_{1,0},u_{1,-1}) - P^{(k)}(u_{0,1},u_{0,0},u_{0,-1}) ,
\eea
\bea\label{a23}
 &&p^{(k)}_{0,-1} - p^{(k)}_{-1,-1} = q^{(k)}_{-1,0} - q^{(k)}_{-1,-1} = \\ \nonumber
 && \quad P^{(k)}(u_{0,1},u_{0,0},u_{0,-1}) - P^{(k)}(u_{-1,1},u_{-1,0},u_{-1,-1}) .
\eea
We can introduce the operators
\begin{equation}\label{a24}
\hat A = \pa_{u_{0,0}} - \frac{f^{(1,1)}_{u_{0,0}}}{f^{(1,1)}_{u_{0,1}}} \pa_{u_{0,1}} -
\frac{f^{(1,-1)}_{u_{0,0}}}{f^{(1,-1)}_{u_{0,-1}}} \pa_{u_{0,-1}} ,
\end{equation}
\begin{equation}\label{a25}
\hat B = \pa_{u_{0,0}} - \frac{f^{(-1,1)}_{u_{0,0}}}{f^{(-1,1)}_{u_{0,1}}} \pa_{u_{0,1}} -
\frac{f^{(-1,-1)}_{u_{0,0}}}{f^{(-1,-1)}_{u_{0,-1}}} \pa_{u_{0,-1}} ,
\end{equation}
such that $\hat A p^{(k)}_{1,-1} = 0$ and $\hat B p^{(k)}_{-1,-1} = 0$. Then we are led to the system
\begin{equation}\label{a26}
\hat A p^{(k)}_{0,-1} = \hat r^{(k,1)} , \quad \hat B p^{(k)}_{0,-1} = \hat r^{(k,2)} ,
\quad [\hat A,\hat B] p^{(k)}_{0,-1} = \hat r^{(k,3)}
\end{equation}
for the function $p^{(k)}_{0,-1}$ depending on $u_{0,1},\,u_{0,0},\,u_{0,-1}$, where $\hat r^{(k,l)}$ are known functions expressed in terms of $f^{(i,j)}$. \qed
\bigskip

After we have solved (\ref{a9}--\ref{a13}) we can construct a generalized symmetry. When both systems (\ref{a21}, \ref{a21a}) and (\ref{a26}) are non-degenerate, we find $\Phi$ and $\Psi$, given by (\ref{a15}), up to at most four arbitrary constants. Two of them may be specified by the consistency conditions (\ref{a14*}) while the remaining constants are specified, using the compatibility condition (\ref{a7}), together with the function $\nu$.
In practice we always look for symmetries of the form (\ref{a5}). Such symmetries are defined uniquely up to  multiple factors and the addition of functions of the form $\nu(u_{0,0})$ corresponding to the right hand side of point symmetries $u_{0,0,\tau} = \nu(u_{0,0})$.   We write down only the generalized symmetry as we are not interested in the point symmetries.

If one of the systems (\ref{a21}, \ref{a21a}) and (\ref{a26}) is degenerate, then $\Phi$ and $\Psi$ in the right hand side of a generalized symmetry (\ref{a15}) may be found up to some arbitrary functions.  Those arbitrary functions must be specified using the compatibility condition (\ref{a7}).
However, in almost all degenerate examples considered below, (\ref{a4}) turns out to be trivial and  it can be rewritten in one of the following four forms:
\begin{equation}\label{a27}
 (T_1 \pm 1) w(u_{0,0},u_{0,1}) = 0 , \qquad (T_2 \pm 1) w(u_{1,0},u_{0,0}) = 0 .
\end{equation}
Eqs. (\ref{a27}) can be integrated once and give equations depending on a reduced number of lattice variables.


In the next Section we test  equations which depend on arbitrary constants and thus we solve some simple classification problem. We look for such particular cases that satisfy our integrability test and are not of  Klein type (\ref{b12}) or not  transformable into Klein type equations by $n,m$-dependent M\"obius transformations.

\section{Examples}\label{sec3}

Here we apply the test to a number of nonlinear nontrivial partial difference equations introduced by various authors using different approaches to prove their integrability \cite{ht94,hyv10,ly09U,lEA,hiv07,abs09}. It should be stressed that all equations below are affine linear, i.e. they can be written in the form (\ref{a1}), (\ref{a3}), where $\pa^2 \mathcal E/\pa u_{n,m}^2 = 0$ for all four variables.

\paragraph{Example 1.}
This will be a simple illustrative example discussed in all details. We consider the equation
\begin{equation}\label{a28}
(u_{1,0} +1) (u_{0,0} -1) = (u_{1,1} -1) (u_{0,1} +1) .
\end{equation}
Up to a  rotation (change of axes)  (\ref{a28}) is equivalent to the equations presented in \cite{ht95} and \cite{nc95}.  In \cite{ly09U,ly09,hyv10}  its  $L-A$ pairs and some conservation laws are presented. Two generalized symmetries of the form (\ref{a5}) have been constructed in \cite{ly09}. So (\ref{a28}) satisfies our integrability test, but nevertheless, it is instructive to try out the test with this equation.

The study of this equation splits into two different steps.

{\bf Step 1.} Let us consider the integrability condition (\ref{a9}) with $k=1$. The corresponding system (\ref{a21}, \ref{a21a}) reads:
\[\begin{array}{l}
 q_{u_{0,0}} - \frac{u_{1,0} + 1}{u_{0,0} - 1} q_{u_{1,0}} - \frac{u_{-1,0} - 1}{u_{0,0} + 1} q_{u_{-1,0}}
 = \frac{2 u_{0,0}}{1 - u_{0,0}^2} , \\
 q_{u_{0,0}} - \frac{u_{1,0} - 1}{u_{0,0} + 1} q_{u_{1,0}} - \frac{u_{-1,0} + 1}{u_{0,0} - 1} q_{u_{-1,0}}
 = \frac{2 u_{0,0}}{1 - u_{0,0}^2} , \\
 (u_{1,0} u_{0,0} + 1) q_{u_{1,0}} - (u_{-1,0} u_{0,0} + 1) q_{u_{-1,0}} = 0 ,
\end{array}\]
where $q = q^{(1)}_{-1,0}$ and by the index we denote the argument of the derivative. This system is non-degenerate, and its solution is
\bea \nonumber
q_{u_{1,0}} = q_{u_{-1,0}} =0,\quad q_{u_{0,0}} = \frac{2 u_{0,0}}{1 - u_{0,0}^2}.
\eea
Hence $q^{(1)}_{0,0} = - \log (u_{1,0}^2 - 1) + c_1$, where $c_1$ is an arbitrary constant.  $q^{(1)}_{0,0}$ together with $p^{(1)}_{0,0} = \log \frac{u_{0,0} - 1}{u_{0,1} + 1}$ satisfy the relation (\ref{a9}), and provide a conservation law for (\ref{a28}).

Eq. (\ref{a9}) with $k=2$ can be solved in a simpler way. As
$p^{(2)}_{0,0} = - p^{(1)}_{0,0} + \log(-1)$, then  the solution of (\ref{a9}) with $k=2$ is given by  $q^{(2)}_{0,0} = - q^{(1)}_{0,0} + c_2$, with $c_2$ another arbitrary constant. As the corresponding system (\ref{a21}, \ref{a21a}) is non-degenerate, there is no other solution.

Now we look for $\Phi$, the r.h.s. of the symmetry in (\ref{a5}). As follows from  (\ref{a14}), a candidate for such a symmetry is given by:
\[
u_{0,0,t_1} = (u_{0,0}^2 - 1) (\alpha u_{1,0} + \beta u_{-1,0}) + \nu(u_{0,0}) .
\]
Here $\alpha$, $\beta$ are nonzero arbitrary constants, and $\nu(u_{0,0})$ is an arbitrary function of its argument. Rescaling $t_1$, we can set in all generality $\alpha = 1$. Substituting  this  into the compatibility condition (\ref{a7}) we get $\beta = -1$, $\nu(u_{0,0}) \equiv 0$. This symmetry is nothing but the well-known modified Volterra equation \cite{y06}:
\begin{equation}\label{sym1}
u_{0,0,t_1} = (u_{0,0}^2 - 1) (u_{1,0} - u_{-1,0}) .
\end{equation}

{\bf Step 2.} Let us consider the integrability conditions (\ref{a9}) with $k=3,4$. In this case the corresponding system (\ref{a26}) is degenerate. So we have to modify the procedure presented in the previous section and applied up above in the case when $k=1,2$. For $k=3$  this system reads:
\[\begin{array}{l}
 (u_{0,1} + u_{0,0}) p_{u_{0,1}} - (u_{0,0} + u_{0,-1}) p_{u_{0,-1}} = 2 , \\
 (u_{0,0}^2 - 1) p_{u_{0,0}} + (u_{0,1} u_{0,0} + 1) p_{u_{0,1}} + (u_{0,0} u_{0,-1} + 1) p_{u_{0,-1}} = 0 ,
\end{array}\]
where $p = p^{(3)}_{0,-1}$. Its general solution is
\bea \label{o1}
p = \Omega(\omega) + \log \frac{u_{0,1} + u_{0,0}}{u_{0,0} + u_{0,-1}} , \quad \mbox{with} \quad
\omega = \frac{u_{0,0}^2 - 1}{(u_{0,1} + u_{0,0})(u_{0,0} + u_{0,-1})} .
\eea
The integrability condition (\ref{a9}) with $k=3$ is satisfied iff $\Omega(\omega) = \gamma - \log \omega$, where $\gamma$ is an arbitrary constant. The case $k=4$ is quite similar, and we easily find the second generalized symmetry of (\ref{a28}):
\begin{equation}\label{sym2}
u_{0,0,t_2} =
(u_{0,0}^2 -1) \left(\frac1{u_{0,1}+u_{0,0}} - \frac1{u_{0,0}+u_{0,-1}}\right) .
\end{equation}

Step 2 of this example is not standard. In all the following  discrete equations,  either the systems (\ref{a21}, \ref{a21a}) and (\ref{a26}) are non-degenerate or the equations themselves are trivial. As a result we have proved the following statement:

\begin{theorem}\label{te2}
Eq. (\ref{a28}) satisfies the generalized symmetry test and possesses the symmetries (\ref{sym1}) and (\ref{sym2}).
\end{theorem}

\paragraph{Example 2.}
Let us consider a known equation closely related to  (\ref{a28}) and studied in
\cite{ht94,mkno97,nah09,hyv10}
\begin{equation}\label{b1}
u_{1,1} (u_{0,1} + c) (u_{1,0} - 1) = u_{0,0} (u_{1,0} + c) (u_{0,1} - 1) ,
\end{equation}
where $c \ne -1,0$. In particular, its $L-A$ pair can be found in \cite{nah09}. If $c = -1$, it is trivial. If $c=0$, using the point transformation
\begin{equation}\label{b2}
u_{n,m} = \frac2{1 - \hat u_{n,m}} ,
\end{equation}
we can reduce it to (\ref{a28}) for $\hat u_{n,m}$. So, (\ref{b1}) generalizes (\ref{a28}). Thus, it will not be surprising that this  equation satisfies our test. The calculation is quite similar to the one shown in Example 1, Step 1. We easily find two generalized symmetries of the form (\ref{a5}):
\bea \label{sym1*}
 \frac{u_{0,0,t_1}}{u_{0,0} (u_{0,0} - 1)} &=& (T_1-1)
   \frac1{u_{0,0} u_{-1,0} + c (u_{0,0} + u_{-1,0} - 1)} , \\ \label{sym2*}
    \frac{u_{0,0,t_2}}{u_{0,0} (u_{0,0} + c)} &=& (T_2-1)
   \frac1{u_{0,0} u_{0,-1} - (u_{0,0} + u_{0,-1} + c)} .
\eea
As a result we get:

\begin{theorem}\label{te3}
Eq. (\ref{b1}) satisfies the generalized symmetry test and possesses the symmetries (\ref{sym1*}) and (\ref{sym2*}).
\end{theorem}

\paragraph{Example 3.}
A further example is given by the equation:
\begin{equation}\label{b3}
u_{1,1} u_{0,0} (u_{1,0} - 1) (u_{0,1} + 1) +  (u_{1,0} + 1) (u_{0,1} - 1) = 0
\end{equation}
taken from ref. \cite{hyv10}. It is  an equation which possesses 5 non--autonomous conservation laws of the form:
\begin{equation}\label{b4}
(T_1-1) p_{n,m} (u_{n,m}, u_{n,m+1}) = (T_2-1) q_{n,m} (u_{n,m}, u_{n+1,m}) ,
\end{equation}
where $p_{n,m}$, $q_{n,m}$ depend explicitly on the discrete variables $n,m$.    In \cite{hyv10} the authors also calculated  the algebraic entropy  for (\ref{b3}), and  demonstrated in this way that the equation should be integrable.

This example does not satisfy our test. The system (\ref{a21}, \ref{a21a}), corresponding to the first of the integrability conditions (\ref{a9}), is non--degenerate, and we find from it $q_{0,0}^{(1)}$ in a unique way. However this function does not satisfy the condition (\ref{a9}). The same is true for all 4 integrability conditions. This means that all 4 assumptions of Theorem \ref{te} are not satisfied.

\begin{theorem}\label{te4}
Eq. (\ref{b3}) does not satisfy any of 4 integrability conditions (\ref{a9})-(\ref{a13}). This equation cannot have an autonomous nontrivial generalized symmetry of the form (\ref{a2}).
\end{theorem}

Eq. (\ref{b3}) might have, however, a non-autonomous generalized symmetry. The extension of the method to non--autonomous generalized symmetries for partial difference equations  is an open problem which is left for  future work.

\paragraph{Example 4.}
Let us consider the equation
\begin{equation}\label{b5}
(1 + u_{0,0} u_{1,0}) (\nu u_{1,1} + u_{0,1})
    = (1 + u_{0,1} u_{1,1}) (\nu u_{0,0} + u_{1,0}) ,
\end{equation}
where the constant $\nu$ is such that $\nu^2 \ne 1$. When $\nu = \pm 1$ the equations are trivial, as they are equivalent to  (\ref{a27}). Eq. (\ref{b5}) has been obtained in \cite{ly09U} by combining Miura type transformations relating differential difference equations of the Volterra type. In \cite{rgsw09} Miura type transformations have been found relating this equation to integrable equations of the form (\ref{a1}). Eq. (\ref{b5}) satisfies our test, and we find 2 generalized symmetries:
\begin{equation}\label{b6}
u_{0,0,t_1} =
 \frac{ (u_{0,0}^2 - \nu) (\nu u_{0,0}^2 - 1) }{u_{0,0}}
 \left( \frac1{u_{1,0} u_{0,0} + 1} - \frac1{u_{0,0} u_{-1,0} + 1} \right) ,
\end{equation}
\begin{equation}\label{b7}
u_{0,0,t_2} =
 \frac{ (u_{0,0}^2 - \nu) (\nu u_{0,0}^2 - 1) }{u_{0,0}}
 \left( \frac1{u_{0,1} u_{0,0} - 1} - \frac1{u_{0,0} u_{0,-1} - 1} \right) .
\end{equation}
In the particular case $\nu = 0$ (\ref{b5}) reduces to
\begin{equation}\label{b8}
u_{1,1} - u_{0,0} = \frac1{u_{1,0}} - \frac1{u_{0,1}} ,
\end{equation}
and (\ref{b6}, \ref{b7}) to its generalized symmetries. Eq. (\ref{b8}), up to point transformations, can be found in \cite{ht95,hyv10,grp91}. As a result of this example we can state the following theorem:

\begin{theorem}\label{te5}
Eq. (\ref{b5}) satisfies the generalized symmetry test and possesses the symmetries (\ref{b6}) and (\ref{b7}).
\end{theorem}

\paragraph{Example 5.}
A further interesting example is provided by the equation \cite{lEA}
\bea \label{iea}
&& \quad 2 (u_{0,0} + u_{1,1}) + u_{1,0} + u_{0,1} + \\    \nonumber
&& \quad  \gamma ( 4 u_{0,0} u_{1,1} + 2 u_{1,0} u_{0,1} + 3 (u_{0,0} + u_{1,1}) (u_{1,0} + u_{0,1}) ) + \\  \nonumber
&& \quad
 (\xi_2 + \xi_4) u_{0,0} u_{1,1} (u_{1,0} + u_{0,1}) + (\xi_2 - \xi_4) u_{1,0} u_{0,1} (u_{0,0} + u_{1,1}) + \\  \nonumber
&& \quad
 \zeta u_{0,0} u_{1,1} u_{1,0} u_{0,1} = 0 ,
\eea
where $\gamma$, $\xi_2$, $\xi_4$ and $\zeta$ are constant coefficients. This equation is obtained as a subclass of the most general multilinear dispersive equation on the square lattice, $\mathcal Q_+$, whose linear part is a linear combination with arbitrary coefficients of $u_{0,0}+u_{1,1}$ and $u_{1,0}+u_{0,1}$. Eq. (\ref{iea}) is contained in the intersection of 5 of the 6 classes of equations belonging to $\mathcal Q_+$ which are reduced to an integrable Nonlinear Schr\"odinger Equation under a multiple scale reduction.

Using the transformation $u_{n,m} = 1/(\hat u_{n,m} - \gamma)$ and redefining the original constants entered in (\ref{iea}):
\[
\alpha = \xi_2 + \xi_4 - 5 \gamma^2 , \qquad
\beta = \xi_2 - \xi_4 - 4 \gamma^2 , \qquad
\delta =  \zeta + 12 \gamma^3 - 4 \gamma \xi_2 ,
\]
we obtain for $\hat u_{n,m}$ a simpler equation depending on just 3  free parameters:
\begin{equation}\label{b9}
(\hat u_{0,0} \hat u_{1,1} + \alpha) (\hat u_{1,0} +\hat  u_{0,1}) + (2 \hat u_{1,0} \hat u_{0,1} + \beta) (\hat u_{0,0} + \hat u_{1,1}) + \delta = 0 .
\end{equation}
If the three parameters are null,  (\ref{b9}) is  a linear equation in $\tilde u_{0,0} = 1/{\hat u_{0,0}}$, and thus trivially integrable.

For  (\ref{b9}) the test is more complicate, as the system (\ref{a21}, \ref{a21a}) depends on the additional variables $ \hat u_{0,1}$, $\hat u_{0,-1}$.  It is written as a polynomial system and setting to zero the coefficients  of the different powers of $ \hat u_{0,1}$ and $\hat u_{0,-1}$ we obtain a simpler system of  equations for $q_{-1,0}^{(k)}$. The same is also true in the case of  the system (\ref{a26}).

Eq. (\ref{b9}) is  a simple classification problem, as it depends on three arbitrary constants, and we search for all integrable cases, if  any, contained in it. By looking at its generalized symmetries we find 2 integrable non-linearizable cases:
\begin{enumerate}
\item $\alpha = 2 \beta \ne 0$ and $\delta = 0$, i.e. $\xi_2 =3 \xi_4 + 3 \gamma^2$, $\zeta = 12 \gamma \xi_4$;
\item $\beta = 2 \alpha \ne 0$ and $\delta = 0$, i.e. $\xi_2 =  6 \gamma^2 - 3 \xi_4$, $\zeta = 12 \gamma (\gamma^2 - \xi_4)$.
\end{enumerate}
In Case 1, using the transformation $\hat u_{n,m} = u_{n,m} (-1)^m \beta^{1/2}$, we obtain (\ref{b5}) with $\nu = 1/2$.
In Case 2 we can always choose  $\alpha = 1$ and the equation reads
\begin{equation}\label{b10}
(u_{0,0} u_{1,1} + 1) (u_{1,0} + u_{0,1}) + 2 (u_{1,0} u_{0,1} + 1) (u_{0,0} + u_{1,1})  = 0 .
\end{equation}
By applying the procedure presented  in the previous section we find the symmetries
\begin{equation}\label{b11}
u_{0,0,t_1} = (u_{0,0}^2 - 1) \frac{u_{1,0} - u_{-1,0}}{u_{1,0} u_{-1,0} - 1} , \qquad
u_{0,0,t_2} = (u_{0,0}^2 - 1) \frac{u_{0,1} - u_{0,-1}}{u_{0,1} u_{0,-1} - 1} .
\end{equation}
This last example (\ref{b10}) seems to be a new integrable model. More comments on this will be presented  in Section \ref{sec4}. This result can be formulated as the following theorem:

\begin{theorem}\label{te6}
There are 2 nontrivial cases when  (\ref{b9}) satisfies the generalized symmetry test. The first one is given by the relations $\alpha = 2 \beta \ne 0$, $\delta = 0$, and the equation is transformed into  (\ref{b5}). In the second case, an equation can be written as  (\ref{b10}) which possesses the symmetries (\ref{b11}).
\end{theorem}

\paragraph{Example 6.}
This example is also an equation with arbitrary constant coefficients, obtained  by Hietarinta and Viallet \cite{hiv07} as an equation with good factorization properties and considered to be an equation worth further studying:
\begin{equation}\label{b14}
(u_{0,0} - u_{0,1}) (u_{1,0} - u_{1,1}) + (u_{0,0} - u_{1,1}) r_4 + (u_{0,1} - u_{1,0}) r_3 + r = 0 .
\end{equation}
The authors claim that (\ref{b14}) is integrable for all values of the coefficients, as it has a quadratic growth of the iterations in the calculation of its algebraic entropy.

  Here we see that if  $r_4=r_3=r=0$, the equation is trivial.
So we consider only those cases when the triple of parameters $r_4, \, r_3$ and $r$ is different from zero.

Let $r_4+r_3=\nu=0$ in  (\ref{b14}). We apply an $n,m$-dependent point transformation $u_{n,m} = \hat u_{n,m} + (n+m) r_4$ and obtain for $\hat u_{n,m}$ the equation
\[
(u_{0,0} - u_{0,1}) (u_{1,0} - u_{1,1}) + r - r_4^2 = 0
\]
of Klein type, more precisely, a particular case of the $Q_V$ equation. However, it is obviously trivial whenever  $r = r_4^2$. If $r \ne r_4^2$, we can rewrite it as
\[
(T_1 + 1) [ \log(u_{0,0} - u_{0,1}) - \frac12 \log(r_4^2 - r) ] = 0 ,
\]
i.e. the equation is trivial in this case too.

The other possible case is when $ r_4+r_3=\nu  \ne 0$. By the transformation $u_{n,m} = \nu \hat u_{n,m}$ we get the following  two parameter equation
\begin{equation}\label{b15}
(u_{0,0} - u_{0,1} + a) (u_{1,0} - u_{1,1} + a) + u_{0,1} - u_{1,0} + b = 0,
\end{equation}
where
\[
r_4 = a \nu , \qquad r_3 = (1-a) \nu , \qquad r = (b + a^2) \nu^2.
\]
This equation has 2 generalized symmetries which we can construct using our procedure:
\begin{equation}\label{b16}
u_{0,0,t_1} = (u_{1,0} - u_{0,0} - a - b) (u_{0,0} - u_{-1,0} - a - b) ,
\end{equation}
\begin{equation}\label{b17}
u_{0,0,t_2} = \frac{(1 - y_{0,0}) (1 - y_{0,-1})}{y_{0,0} + y_{0,-1}} + 1 ,  \quad
y_{n,m} = 2 (u_{n,m+1} - u_{n,m}) - 2 a + 1,
\end{equation}
showing its integrability. This result can be formulated as:

\begin{theorem}\label{te7}
In the case $r_4+r_3=0$, (\ref{b14}) is equivalent to a trivial equation. In the case $r_4+r_3 \ne 0$, it can be rewritten in the form (\ref{b15}). Eq. (\ref{b15}) satisfies the generalized symmetry test and possesses the symmetries (\ref{b16}) and (\ref{b17}).
\end{theorem}

\paragraph{Example 7.}
The next example is also taken from \cite{hiv07}:
\begin{equation}\label{b18}
   u_{0,0} u_{0,1} c_5 + u_{1,0} u_{1,1} c_6
+  u_{0,0} u_{1,0} c_1 + u_{0,1} u_{1,1} c_3
+ (u_{0,0} u_{1,1} + u_{1,0} u_{0,1}) c_2 = 0 .
\end{equation}
This equation is proven to be integrable for all values of constants $c_i$ by checking its algebraic entropy.
Also in this case we  have a kind of classification problem once we  exclude, up to some simple transformations, all Klein type and trivial subequations.

Let us observe at first that if $c_5 = c_6$ and $c_1 = c_3$,   (\ref{b18}) is of Klein type, and if moreover $c_1=c_5=0$, it is trivial. We can construct some point transformations which leave (\ref{b18}) invariant, but transform the coefficients among themselves. By the transformation
\bea
u_{n,m} = \hat u_{m,n} ,  \label{b19}	
\eea
$c_5 \leftrightarrow c_1$, $c_6 \leftrightarrow c_3$, and by the transformation
 \bea
u_{n,m} = 1/\hat u_{n,m} , \label{b20}
\eea
$c_5 \leftrightarrow c_6$, $c_1 \leftrightarrow c_3$. In both cases $c_2$ remains unchanged.  Moreover, the $n,m$ dependent transformation
\begin{equation}\label{b21}
	u_{n,m} = \hat u_{n,m} \kappa_1^n \kappa_2^m , \qquad \kappa_i \ne 0, \, i=1,2,
\end{equation}
leaves the equation invariant  with the following transformation of the coefficients:
\[
\hat c_5 = c_5/\kappa_1 , \quad \hat c_6 = c_6 \kappa_1 , \quad \hat c_1 = c_1/\kappa_2 , \quad
\hat c_3 = c_3 \kappa_2 , \quad \hat c_2 = c_2 .
\]

So, if at least one of the coefficients $c_i$ ($i \ne 2$) is different from  zero,  using the transformations (\ref{b19}) and (\ref{b20}), we can make $c_5 \ne 0$.
Let us assume that also $c_6 \ne 0$. If either $c_1$ or $c_3$ is equal to zero  then, using the transformations (\ref{b19}) and (\ref{b20}), we can make   $c_6 = 0$. If both  $c_1$ and $c_3$ are either zero or different from zero, using the transformation (\ref{b21}), we can make $c_1=c_3$ and $c_5=c_6$, i.e. we obtain a Klein type equation. So the only possible remaining  case is when $c_5 \ne 0$, $c_6 = 0$ and without loss of generality we can set
\begin{equation}\label{b22}
	c_5 = 1 , \quad c_6 = 0 .
\end{equation}
The non-degeneracy conditions (\ref{a3}) give 2 restrictions: $c_2 \ne 0$ or $c_2 = 0$, $c_1 c_3 \ne 0$. In these 2 cases, the equation can be nontrivially rewritten in the form of (\ref{a4}). If $c_2 \ne 0$ and $c_1=c_3=0$,  (\ref{b18}) is trivial, as it is equivalent to
\[
(T_2+1) \left( c_2 \frac{u_{1,0}}{u_{0,0}} + \frac12 \right) = 0.
\]
 So, at the end  we get  2 admissible cases:
\begin{eqnarray}
 c_2 = 0: \ \qquad c_1 c_3 \ne 0 , \label{b23}	\\
 c_2 \ne 0: \quad c_1 \ \hbox{or} \ c_3 \ne 0 . \label{b24}
\end{eqnarray}

Any equation (\ref{b18}), (\ref{b22}) satisfying conditions (\ref{b23}), (\ref{b24}) possesses 2 generalized symmetries. The first symmetry depends on the number $c_1 c_3 - c_2^2$. If
\begin{equation}\label{b25}
	c_1 c_3 - c_2^2 \ne 0 ,
\end{equation}
the condition (\ref{b23}) is satisfied automatically.  The symmetry reads:
\begin{equation}\label{b26}
	u_{0,0,t_1} = (u_{1,0} - c u_{0,0}) \left( \frac{u_{0,0}}{u_{-1,0}} - c \right) , \qquad
	c = \frac{c_2}{c_1 c_3 - c_2^2} .
\end{equation}
In the case when $c_1 c_3 = c_2^2$, as $c_2 \ne 0$ due to condition (\ref{b23}), the  condition (\ref{b24}) is satisfied automatically. In this case
\begin{equation}\label{b27}
	c_1 c_3 = c_2^2 \ne 0 ,
\end{equation}
and the symmetry reads:
\begin{equation}\label{b28}
	u_{0,0,t_1} = u_{1,0} + \frac{u_{0,0}^2}{u_{-1,0}} .
\end{equation}

The form of the second symmetry depends on the number $c_1 c_3$. If
\begin{equation}\label{b29}
	c_1 c_3 \ne 0 ,
\end{equation}
then both non-degeneracy conditions are satisfied, and we have the symmetry
\begin{equation}\label{b30}
	u_{0,0,t_2} = \frac{c_2 c_3 c_1 (u_{0,1} u_{0,-1} + u_{0,0}^2) +
	\frac12 (c_2^2 + c_3 c_1) u_{0,0} (u_{0,1} c_3 + u_{0,-1} c_1)} {u_{0,1} c_3 - u_{0,-1} c_1} .
\end{equation}
If $c_1 c_3 = 0$, then we cannot have $c_2 = 0$ due to condition (\ref{b23}). So, $c_2 \ne 0$, and we use condition (\ref{b24}). We have here 2 cases for which both non-degeneracy conditions are satisfied. First of them is
\begin{equation}\label{b31}
	c_3 = 0 , \quad c_1 c_2 \ne 0 ,
\end{equation}
and the corresponding symmetry has the form:
\begin{equation}\label{b32}
	u_{0,0,t_2} = \left( u_{0,1} + u_{0,0} \frac{c_1}{c_2} \right)
	\left( \frac{u_{0,0}}{u_{0,-1}} + \frac{c_1}{c_2} \right) .
\end{equation}
The second case is
\begin{equation}\label{b33}
	c_1 = 0 , \quad c_2 c_3 \ne 0 ,
\end{equation}
and the symmetry reads:
\begin{equation}\label{b34}
	u_{0,0,t_2} = \left( \frac{u_{0,0}}{u_{0,1}} + \frac{c_3}{c_2} \right)
	\left( u_{0,-1} + u_{0,0} \frac{c_3}{c_2} \right) .
\end{equation}

\begin{theorem}\label{te8}
For  (\ref{b18}) we have:
\begin{enumerate}
	\item  If it is not equivalent to a Klein type equation, then it can be rewritten in the form (\ref{b22}) by using transformations (\ref{b19}), (\ref{b20});
	\item Nontrivial eqs. (\ref{b18}), (\ref{b22}) must satisfy the conditions (\ref{b23}), (\ref{b24});
	\item Eq. (\ref{b18}), (\ref{b22}) with the restriction (\ref{b23}), (\ref{b24}) satisfies the generalized symmetry test for any values of $c_1,c_2,c_3$;
	\item The first symmetry of this equation is of the form (\ref{b26}) in case (\ref{b25}) and of the form (\ref{b28}) in case (\ref{b27});
	\item The second symmetry is of the form (\ref{b30}) in case (\ref{b29}), of the form (\ref{b32}) in case (\ref{b31}) and of the form (\ref{b34}) in case (\ref{b33}).
\end{enumerate}
\end{theorem}

The resulting equations (\ref{b18}), (\ref{b22}) satisfying conditions (\ref{b23}), (\ref{b24}) will be written down in a simpler  explicit form in Section \ref{sec4}. One of these equations, in a slightly different form, can be found in \cite{r10}, where its $L-A$ pair is given. There it is stated that  hierarchies of generalized symmetries and conservation laws exist.

\paragraph{Example 8.}
The last example is taken from an article by Adler, Bobenko and Suris \cite{abs09}, where an extended definition of 3D-consistency is discussed and the so-called deformations of H equations are presented. As an example  let us consider here one of them, namely,
\begin{equation}\label{b35}
 (u_{0,0} - u_{1,1}) (u_{1,0} - u_{0,1}) = (\alpha - \beta) (1 - \epsilon u_{1,0} u_{0,1}) ,
\end{equation}
where $\alpha \ne \beta$ and $\epsilon$ are constants. Eq. (\ref{b35}) is a generalization of the well-known discrete potential KdV or $H_1$ equation which is reobtained when   $\epsilon = 0$.

Let us use the integrability condition (\ref{a9}) with $k=1$ and obtain the system (\ref{a21}, \ref{a21a}). The first equation of this system depends on the additional variable $u_{0,1}$. We rewrite the equation in polynomial form and obtain a 4th degree polynomial in $u_{0,1}$. The coefficients of this polynomial provide us with 5 more  equations for $q_{-1,0}^{(1)}$. Using these equations, we easily obtain as an integrability  condition that $\epsilon = 0$. The other integrability conditions are similar, and none of them is satisfied if $\epsilon \ne 0$.

\begin{theorem}\label{te9}
Eq. (\ref{b35}) with $\epsilon \ne 0$ satisfies none of 4 integrability conditions (\ref{a9})-(\ref{a13}). This equation cannot have an autonomous nontrivial generalized symmetry of the form (\ref{a2}).
\end{theorem}

The result is not surprising, as  (\ref{b35}) is  3D-consistent on the so-called black-white lattice. This means that  to check 3D-consistency we have to use (\ref{b35}) together with another equation, i.e.  (\ref{b35}) is conditionally 3D-consistent. Generalized symmetries might exist in a similar indirect sense when we consider the complete 3D--consistent system. The following $n,m$-dependent equation
\begin{equation}\label{b36}
\begin{array}{l}
 (u_{n,m} - u_{n+1,m+1}) (u_{n+1,m} - u_{n,m+1}) - (\alpha - \beta) + \\
 \epsilon (\alpha - \beta) ( \frac{1+(-1)^{n+m}}2 u_{n+1,m} u_{n,m+1} + \frac{1-(-1)^{n+m}}2 u_{n,m} u_{n+1,m+1} ) = 0
\end{array}
\end{equation}
is obtained in \cite{xp09} instead of (\ref{b35}). Eq. (\ref{b35}) is obtained when $n+m$ is even, while if $n+m$ is odd we have a different equation. An $n,m$-dependent $L-A$ pair and $n,m$-dependent generalized symmetries have been constructed in \cite{xp09} for (\ref{b36}). Such $n,m$-dependent generalized symmetries could be possibly constructed, starting from its $L-A$ pair.

\section{General picture}\label{sec4}

We have applied our test to a number of discrete equations on the square lattice and have constructed generalized symmetries for some of them. Such equations have automatically a few simple conservation laws. Here we collect together all these equations satisfying the test in order to discuss and compare them.

Discrete-differential equations of the Volterra type
\begin{equation}\label{c1}
	u_{k,t} = \phi (u_{k+1}, u_k, u_{k-1})
\end{equation}
play an important role in this discussion. The main representative of this class is the well-known Volterra equation.
A complete list of integrable equations of the Volterra type has been obtained, using the generalized symmetry method, in  \cite{y83}, see the review \cite{y06} for details. As (\ref{c1}) are autonomous, they  will be written down below at $k=0$.

The most interesting example of an equation of the class (\ref{c1}), apart from the Volterra equation, is the equation:
\begin{equation}\label{c2}
	\dot u_0 = \frac{r}{u_1 - u_{-1}} - \frac12 \frac{\pa r}{\pa u_1},
\end{equation}
where $r = r(u_1, u_0)$ is an arbitrary bi-quadratic and symmetric polynomial in its 2 arguments with 6 constant coefficients. This is an integrable discretization found by R. Yamilov in \cite{y83}, from now on  abbreviated as the YdKN equation, of the well-known Krichever-Novikov equation. Two different representations of(\ref{c2}) can be found in \cite{y83} and \cite{y06}. The YdKN equation can be obtained as the continuous limit of the $Q_4$ equation \cite{as04}. This limit preserve the 3D-- consistency condition and thus the YdKN equation is a symmetry of the $Q_4$ \footnote{Communication of the Referee.}

It has been observed in \cite{lpsy08} that generalized symmetries (\ref{a5}) of any equation of the ABS list are of the form (\ref{c2}), i.e. they are subcases of the YdKN  equation. In  \cite{lpsy08}, it has been explained  that equations of the ABS list can be interpreted as  auto--B\"acklund transformations of their symmetries, i.e. of YdKN type equations. Moreover  particular cases of the generalized symmetries of the YdKN equation are generalized symmetries of the equations of the ABS list.

All generalized symmetries mentioned in previous Section can be identified, up to point transformations $u_{n,m} = \omega (\hat u_{n,m})$, with an equation of the complete list of integrable equations of the Volterra type (\ref{c1}), presented  in \cite{y06}. Following \cite{lpsy08}, we are going to use here this relation between Volterra type equations and discrete equations on the square lattice. Let us present here some of the reasons why it is convenient to use this connection:

\begin{enumerate}

	\item One can always interpret, as it has been done in \cite{lpsy08}, discrete equations on the square  satisfying our test as B\"acklund transformations for their symmetries. B\"acklund transformations of integrable equations are integrable equations too, as they are characterized by a Lax pair as the nonlinear equations themselves.
	
	\item If we know hierarchy of generalized symmetries for a Volterra type equation, we automatically obtain hierarchy of generalized symmetries for its B\"acklund transformation.
	
	\item Classification of integrable Volterra type equations up to Miura type transformations can be found in \cite{y06}. This  suggests the Miura type transformations relating different discrete equations on the square lattice.
		
	\item Is is not so easy to check whether 2 discrete equations are different up to M\"obius transformations. A relatively easy way to do it is by comparing their symmetries\footnote{An alternative way to compare discrete equations is by use of the invariants of M\"obius transformations \cite{abs09}.}, i.e. equations (\ref{c1}), as  generalized symmetries for these equations have been constructed in unique way up to point symmetries. We can even  do that for $n,m$-dependent M\"obius transformations.

\end{enumerate}

In \cite{x09} it is shown that generalized symmetries (\ref{a5}) for Klein type equations are always given by particular cases of the YdKN equation. The  two symmetries may be different, but both are particular cases of the YdKN equation. As a result we have the following picture:
\[
\hbox{YdKN equation}: \quad \hbox{Klein type equations} \  \equiv \ Q_V \ \hbox{equation} \ \supset \ \hbox{ABS list}
\]
i.e. Klein type equations are auto--B\"acklund transformations for particular cases of the YdKN equation.
 So the Klein type equations are  essentially equivalent to the $Q_V$ equation, see Appendix A,
which includes in its turn the ABS list \cite{x09}. This picture is true only in the autonomous case. In general, the ABS equations may be lattice dependent and are not included in $Q_V$.  Moreover, generically $Q_V$ is just equivalent to $Q_4$ up to a M\"obius transformation \footnote{Communication of the Referee}. $Q_4$, due to its special parametrization of the coefficients, possesses a Lax pair characterized by copies of the Lax operator while this is not the case for $Q_V$. The ABS list includes a number of well-known nonlinear partial difference equations, see a review in \cite{abs03}. A hierarchy of generalized symmetries for the YdKN equation has been constructed, using a master symmetry  \cite{asy00}, see also a detailed discussion in \cite{lpsy08}. In this way we obtain generalized symmetries for all Klein type equations. An alternative way for constructing symmetries for Klein type equations can be found in \cite{x09}. Many  subcases of the YdKN equation can be transformed, using Miura type transformations, into the Volterra or Toda lattice equations \cite{y06}.

\begin{table}
\caption{The nontrivial non--Klein type partial difference equations analyzed in  Section \ref{sec3} and their generalized symmetries written in the form (\ref{a5}). In the last column we give the number of an example of Section \ref{sec3} where corresponding equation is considered.}
 \begin{center}
\begin{tabular}{|c|c|c|c|}
 \hline
 Eq. & Difference Equations & $ \Phi (u_{1,0}, u_{0,0}, u_{-1,0}) $ & Ex. \\
 N   &                      & $ \Psi (u_{0,1}, u_{0,0}, u_{0,-1}) $ & N   \\\hline \hline
T1 & ${\scriptstyle u_{1,1} (u_{0,1} + c) (u_{1,0} - 1) = u_{0,0} (u_{1,0} + c) (u_{0,1} - 1)  }$ &
${\scriptstyle  \frac{u_{0,0} (u_{0,0} - 1)}{u_{1,0} u_{0,0} + c (u_{1,0} + u_{0,0} - 1)}
 - \frac{u_{0,0} (u_{0,0} - 1)}{u_{0,0} u_{-1,0} + c (u_{0,0} + u_{-1,0} - 1)}}$ & 2 \\
 & ${\scriptstyle c \ne -1}$ & ${\scriptstyle  \frac{u_{0,0} (u_{0,0} + c)}{u_{0,1} u_{0,0} - (u_{0,1} + u_{0,0} + c)}
 - \frac{u_{0,0} (u_{0,0} + c)}{u_{0,0} u_{0,-1} - (u_{0,0} + u_{0,-1} + c)}}$ & \\
  \hline
T2 & ${\scriptstyle (u_{0,0} u_{1,1} - 1) (u_{1,0} + u_{0,1} - \kappa) = (1 - u_{1,0} u_{0,1}) (u_{0,0} + u_{1,1} + \kappa)  }$ & ${\scriptstyle (u_{0,0}^4 + (2-\kappa^2) u_{0,0}^2 + 1)}
   \left( \frac1{u_{1,0} + u_{0,0}}  - \frac1{u_{0,0} + u_{-1,0}} \right) $ & 4 \\
 & ${\scriptstyle \kappa \ne 0}$ & ${\scriptstyle (u_{0,0}^4 + (2-\kappa^2) u_{0,0}^2 + 1) } \left( \frac1{u_{0,1} + u_{0,0}}  - \frac1{u_{0,0} + u_{0,-1}} \right)$ &\\
 \hline
T3 & ${\scriptstyle 3 (u_{0,0} u_{1,1} u_{1,0} u_{0,1} - 1) + u_{0,0} u_{1,1} - u_{1,0} u_{0,1} = 0}$ &
${\scriptstyle  2 u_{0,0} \frac{u_{1,0} - u_{-1,0}}{u_{1,0} + u_{-1,0}}}$ & 5 \\
  &  & ${\scriptstyle  2 u_{0,0} \frac{u_{0,1} - u_{0,-1}}{u_{0,1} + u_{0,-1}}}$ &  \\
 \hline
T4 & ${\scriptstyle (u_{0,0} - u_{0,1} + 1/2) (u_{1,0} - u_{1,1} + 1/2) + u_{0,1} - u_{1,0} = 0}$ & $ {\scriptstyle (u_{1,0} - u_{0,0} - 1/2) (u_{0,0} - u_{-1,0} - 1/2) }$ & 6 \\
 & & ${\scriptstyle  \frac{(u_{0,1} - u_{0,0}) (u_{0,0} - u_{0,-1}) + 1/4}{u_{0,1} - u_{0,-1}}}$ & \\
\hline
T5 & ${\scriptstyle u_{0,0} (u_{1,0} + u_{0,1} + u_{1,1}) + u_{1,0} u_{0,1} = 0}$ & ${\scriptstyle (u_{1,0} + u_{0,0}) \left( \frac{u_{0,0}}{u_{-1,0}} + 1 \right) }$ & 7 \\
 & & ${\scriptstyle (u_{0,1} + u_{0,0}) \left( \frac{u_{0,0}}{u_{0,-1}} + 1 \right)}$ & \\
 \hline
T6 & ${\scriptstyle(u_{1,1} - u_{1,0}) (u_{0,1} - u_{0,0}) + u_{0,1} u_{0,0} = 0}$ & ${\scriptstyle  u_{1,0} + \frac{u_{0,0}^2}{u_{-1,0}} }$ & 7 \\
 & & ${\scriptstyle  \frac{(u_{0,1} - u_{0,0}) (u_{0,0} - u_{0,-1})}{u_{0,1} - u_{0,-1}}}$ & \\
 \hline
T7 & ${\scriptstyle  u_{0,1} u_{1,1} + u_{0,0} (u_{1,0} + u_{0,1}) +
 (u_{0,0} u_{1,1} + u_{1,0} u_{0,1}) c_2 = 0  }$ & ${\scriptstyle \left( u_{1,0} + u_{0,0} \frac{c_2}{c_2^2-1} \right)
	\left( \frac{u_{0,0}}{u_{-1,0}} + \frac{c_2}{c_2^2-1} \right) }$ & 7 \\
	& ${\scriptstyle  c_2^2 \ne 1}$ & ${\scriptstyle \frac{2 c_2 (u_{0,1} u_{0,-1} + u_{0,0}^2) +
	(c_2^2 + 1) u_{0,0} (u_{0,1} + u_{0,-1})} {u_{0,1} - u_{0,-1}}}$ & \\
 \hline
T1* & ${\scriptstyle (u_{1,0} +1) (u_{0,0} -1) = (u_{1,1} -1) (u_{0,1} +1) }$ &
${\scriptstyle (u_{0,0}^2 - 1) (u_{1,0} - u_{-1,0})}$ & 1 \\
 & & ${\scriptstyle (u_{0,0}^2 -1) \left(\frac1{u_{0,1}+u_{0,0}} - \frac1{u_{0,0}+u_{0,-1}}\right) }$ & 2 \\
 \hline
T2* & ${\scriptstyle u_{1,1} - u_{0,0} = \frac1{u_{1,0}} - \frac1{u_{0,1}} }$ & ${\scriptstyle u_{0,0} \left( \frac1{u_{1,0} u_{0,0} + 1} - \frac1{u_{0,0} u_{-1,0} + 1} \right) }$ & 4 \\
 & & ${\scriptstyle u_{0,0} \left( \frac1{u_{0,1} u_{0,0} - 1} - \frac1{u_{0,0} u_{0,-1} - 1} \right) }$ & \\
 \hline
\end{tabular}
\end{center}
\end{table}

Let us write down in Table 1 the nontrivial non--Klein type equations, satisfying the test, together with their generalized symmetries. We present those equations in a simpler or slightly different form convenient for this Section.

Eq. (T1) of Table 1 is nothing else but  (\ref{b1}) of Example 2 with its symmetries. If $c=-1$, it is degenerate. If $c=0$, the transformation (\ref{b2}) gives (\ref{a28}) with both its symmetries, as it is explained in Example 2.

Eq. (T2) of Table 1 is obtained from  (\ref{b5}) of Example 4. If in  (\ref{b5}) we set $\nu = 0$, then we have  (\ref{b8}). By the lattice-dependent point transformation
$u_{n,m} = i \frac{\hat u_{n,m} - 1}{\hat u_{n,m} + 1} (-1)^m$ we transform (\ref{b5}) into  (T2) by defining
$\kappa = 2 \frac{1-\nu}{1+\nu}$. In order to obtain its symmetries  we rescale $t_1$ and $t_2$. The case when $\kappa=0$ is trivial in the sense of  (\ref{a27}). If $\kappa=-2$, the transformation is undefined. However, the symmetries are compatible with the equation for any value of the constant $\kappa$.

Eq. (T3) of Table 1 is obtained from (\ref{b10}) of Example 5 by applying the transformation $u_{n,m} = \frac{\hat u_{n,m} - 1}{\hat u_{n,m} + 1}$.

Eq. (T4) of Table 1 is derived from (\ref{b15}) of Example 6, using lattice-dependent point transformation
\[
u_{n,m} = \hat u_{n,m} + (b+a-1/2) n + (a-1/2) m
\]
which allows $a=1/2$, $b=0$.

Eqs. (T5 -- T7) of Table 1, together with the example
\begin{equation}\label{5*}
u_{0,1} (u_{0,0} + u_{1,1} + u_{1,0}) + u_{0,0} u_{1,1} = 0 ,
\end{equation}
\begin{equation}\label{sym5*}
u_{0,0,t_1} = (u_{1,0} + u_{0,0}) \left( \frac{u_{0,0}}{u_{-1,0}} + 1 \right) , \qquad
u_{0,0,t_2} = \left( \frac{u_{0,0}}{u_{0,1}} + 1 \right) (u_{0,-1} + u_{0,0}) ,
\end{equation}
are obtained from Example 7, i.e. (\ref{b18}, \ref{b22}) satisfying conditions (\ref{b23}, \ref{b24}). We consider all possible cases and remove some constants by using transformations of the form (\ref{b21}). Eq. (\ref{5*}) together with its symmetries (\ref{sym5*}) is transformed into (T5) of Table 1 by the transformation $u_{n,m} = \hat u_{n,-m}$ which is not standard for this paper. For this reason,  (\ref{5*}) is not included in Table 1.

Eqs. (T1*, T2*) of Table 1 are particular cases of (T1,T2), respectively, as it is shown in Examples 2 and 4. However, these equations are interesting and well-known as themselves and are included in Table 1 to provide a more complete picture.

Comparing the generalized symmetries, we can easily show that the main 7 equations of the Table are different. More precisely, we have the following statement:

\begin{theorem}\label{te10}
Up to ($n,m$)--dependent point transformations $u_{n,m} = \omega_{n,m} (\hat u_{n,m})$, (T1--T7) of Table 1 are different from Klein type equations and from each other.
\end{theorem}

Its proof is more or less obvious. We will give below some special comments only in the case of (T1) and (T3) of Table 1.

The second symmetries of (T4, T6, T7) of Table 1 are particular cases of the YdKN equation. In these cases we have no problem to find further generalized symmetries. This result shows that we can obtain for the YdKN equation also auto--B\"acklund transformations which are not equations of  Klein type.

From the point of view of its generalized symmetries, (T3) of Table 1 is close to a Klein type equation. Indeed, using the lattice-dependent point transformation $u_{n,m} = \hat u_{n,m} i^{n+m}$, we get from its symmetries the  equations:
\[
 u_{0,0,t_1} = 2 u_{0,0} \frac{u_{1,0} + u_{-1,0}}{u_{1,0} - u_{-1,0}} , \qquad
 u_{0,0,t_2} = 2 u_{0,0} \frac{u_{0,1} + u_{0,-1}}{u_{0,1} - u_{0,-1}} ,
\]
which are both of the YdKN type as in case of Klein type equations. By such a transformation this equation  becomes, however, explicitly lattice-dependent:
\[
3 (u_{n,m} u_{n+1,m+1} u_{n+1,m} u_{n,m+1} - 1) = (-1)^{n+m} (u_{n,m} u_{n+1,m+1} - u_{n+1,m} u_{n,m+1}) .
\]
So, this equation is not a Klein type equation, but it provides $n,m$-dependent B\"acklund transformation for a YdKN type equation. Generalized symmetries for this equation can be obtained, starting from the YdKN equation, but those symmetries may be explicitly $n,m$-dependent due to the involved  transformation.

Let us consider the following  integrable Volterra type equations (\ref{c1}):
\begin{equation}\label{c3}
	\dot u_0 = (\alpha u_0^2 + \beta u_0 + \gamma) (u_1 - u_{-1}) ,
\end{equation}
\begin{equation}\label{c4}
	\dot u_0 = (\alpha u_0^4 + \beta u_0^2 + \gamma) \left( \frac1{u_1 + u_0} - \frac1{u_0 + u_{-1}} \right) ,
\end{equation}
with $\alpha$, $\beta$ and $\gamma$ constant coefficients. Up to linear point transformations, (\ref{c3}) contains   2 nonlinear equations: the Volterra equation if $\alpha=\gamma=0$, $\beta=1$ and the modified Volterra equation if $\alpha=1$, $\beta=0$. Eq. (\ref{c4}) can be called a twice modified Volterra equation, as there is a Miura type transformation from (\ref{c4}) into the modified Volterra equation \cite{y94}. Generalized symmetries of eqs. (\ref{c3}) can be constructed in many different ways, see e.g. \cite{y06}. Generalized symmetries for (\ref{c4}) can be obtained, using a master symmetry found in \cite{cy95}. The additional equation (T1*) of Table 1 provides us examples of symmetries of both types (\ref{c3}) and (\ref{c4}).

The three-point symmetries of (T2) of Table 1  have the form (\ref{c4}). Also the symmetries of (T1) of Table 1 (in the generic case $c \ne 0$) can be rewritten in the form (\ref{c4}), using M\"obius transformations (\ref{b13}). However, we cannot rewrite them both as  (\ref{c4}), using the same M\"obius transformation. In particular (T1) cannot be written in symmetric form as in the case of (T2). So, (T1) and (T2) of Table 1 are different. Their generalized symmetries can be taken from (\ref{c4}).

All the other generalized symmetries are related to the following integrable Volterra type equations:
\begin{eqnarray}
 \dot u_0 &=& (u_{1} - u_{0} + \delta) (u_{0} - u_{-1} + \delta) , \label{c5}	\\
 \dot u_0 &=& (e^{u_{1} - u_{0}} + \delta) (e^{u_{0} - u_{-1}} + \delta) , \label{c6}	\\
 \dot u_0 &=& e^{u_{1} - u_{0}} + e^{u_{0} - u_{-1}} , \label{c7}	
\end{eqnarray}
where $\delta$ is constant. In fact the first symmetry of (T4) of Table 1 is of form of (\ref{c5}). The first symmetry of (T6) of Table 1 is obtained from  (\ref{c7}) by point transformation $\hat u_k = e^{u_k}$. The other symmetries are obtained from  (\ref{c6}): the both symmetries of (T5) and the first symmetry of (T7) of Table 1 are obtained, using the transformation $\hat u_k = e^{u_k}$.

Eqs. (\ref{c5}--\ref{c7}) are slight modifications of  (\ref{c3}). Indeed, using the transformations
\begin{eqnarray}
 \hat u_0 &=& u_{1} - u_{0} + \delta , \label{c8}	\\
 \hat u_0 &=& e^{u_{1} - u_{0}} + \delta , \label{c9}	\\
 \hat u_0 &=& e^{u_{1} - u_{0}} , \label{c10}	
\end{eqnarray}
respectively, we transform (\ref{c5}--\ref{c7}) into equations of the form (\ref{c3}). As the transformations (\ref{c8}--\ref{c10}) are very simple, we can use these transformations together with the symmetries of (\ref{c3})  to construct generalized symmetries for (\ref{c5}--\ref{c7}).
\section*{Appendix}
\appendix
\section{Klein symmetries and $Q_V$}

Let us consider the most general multi-linear equation (\ref{a1}):
\begin{eqnarray}
 && u_{0,0} u_{1,0} u_{0,1} u_{1,1} k_1 + \nonumber	\\
 && u_{1,0} u_{0,1} u_{1,1} k_2 + u_{0,0} u_{0,1} u_{1,1} k_3 +
    u_{0,0} u_{1,0} u_{1,1} k_4 + u_{0,0} u_{1,0} u_{0,1} k_5 +  \label{u1}	\\
 && u_{0,0} u_{1,0} k_6 + u_{0,0} u_{0,1} k_7 + u_{0,0} u_{1,1} k_8 +
    u_{1,0} u_{0,1} k_9 + u_{1,0} u_{1,1} k_{10} + u_{0,1} u_{1,1} k_{11} +  \nonumber \\
 && u_{0,0} k_{12} + u_{1,0} k_{13} + u_{0,1} k_{14} + u_{1,1} k_{15} + k_{16} = 0 . \nonumber	
\end{eqnarray}
Imposing the discrete symmetries
\begin{equation}\label{u2}
\mathcal E (u_{0,0}, u_{1,0}, u_{0,1}, u_{1,1}) = \mathcal E (u_{1,0}, u_{0,0}, u_{1,1}, u_{0,1})
= \mathcal E (u_{0,1}, u_{1,1}, u_{0,0}, u_{1,0}) ,
\end{equation}
we get the $Q_V$ equation
\begin{eqnarray}
 &&  u_{0,0} u_{1,0} u_{0,1} u_{1,1} k_1 + \nonumber	\\
 && (u_{1,0} u_{0,1} u_{1,1} + u_{0,0} u_{0,1} u_{1,1} +
     u_{0,0} u_{1,0} u_{1,1} + u_{0,0} u_{1,0} u_{0,1}) k_2 +  \label{u3}	\\
 && (u_{0,0} u_{1,0} + u_{0,1} u_{1,1}) k_6 + (u_{0,0} u_{0,1} + u_{1,0} u_{1,1}) k_7 +
    (u_{0,0} u_{1,1} + u_{1,0} u_{0,1}) k_8  +  \nonumber \\
 && (u_{0,0} + u_{1,0} + u_{0,1} + u_{1,1}) k_{12} + k_{16} = 0. \nonumber	
\end{eqnarray}
By direct calculation, we can check that the $Q_V$ equation is invariant under an $n,m$-independent M\"obius transformation.

A Klein type  equation  satisfies the following discrete symmetries
\begin{equation}\label{u4}
\mathcal E (u_{0,0}, u_{1,0}, u_{0,1}, u_{1,1}) = \pi_1 \mathcal E (u_{1,0}, u_{0,0}, u_{1,1}, u_{0,1})
= \pi_2 \mathcal E (u_{0,1}, u_{1,1}, u_{0,0}, u_{1,0}) ,
\end{equation}
where $\pi_1 = \pm 1$, $\pi_2 = \pm 1$. In addition to the equation $Q_V$, we have 3 other possible cases.

If $\pi_1 = 1$ and $\pi_2 = -1$, we obtain:
\begin{eqnarray}
 && (u_{1,0} u_{0,1} u_{1,1} + u_{0,0} u_{0,1} u_{1,1} -
     u_{0,0} u_{1,0} u_{1,1} - u_{0,0} u_{1,0} u_{0,1}) k_2 + \nonumber	\\
 && (u_{0,0} u_{1,0} - u_{0,1} u_{1,1}) k_6 +
    (u_{0,0} + u_{1,0} - u_{0,1} - u_{1,1}) k_{12} = 0 . \label{u5}	
\end{eqnarray}
The case when $\pi_1 = -1$ and  $\pi_2 = 1$ is equivalent to previous one up to transformation $u_{n,m} = \hat u_{m,n}$.

If $\pi_1 = -1$ and $\pi_2 = -1$, we obtain:
\begin{eqnarray}
 && (u_{1,0} u_{0,1} u_{1,1} - u_{0,0} u_{0,1} u_{1,1} -
     u_{0,0} u_{1,0} u_{1,1} + u_{0,0} u_{1,0} u_{0,1}) k_2 + \nonumber	\\
 && (u_{0,0} u_{1,1} - u_{1,0} u_{0,1}) k_8 +
    (u_{0,0} - u_{1,0} - u_{0,1} + u_{1,1}) k_{12} = 0 . \label{u6}	
\end{eqnarray}
Eqs. (\ref{u5}), (\ref{u6}) are invariant under M\"obius transformations, thus showing that this property is valid for all Klein type equations.

Let us consider (\ref{u5}). If $k_2=k_{12}=0$, the equation is degenerate:
\[
(T_2-1) (u_{0,0} u_{1,0} k_6) = 0 .
\]
If either $k_2$ or $k_{12}$ is not zero, we can make $k_2 \ne 0$ by using the transformation $u_{n,m} = 1 / \hat u_{n,m}$. Then, using $u_{n,m} = \hat u_{n,m} + k_6 / (2 k_2)$, we make $k_6=0$. Eq. (\ref{u5}) with $k_6=0$ is reduced by the lattice-dependent transformation $u_{n,m} = \hat u_{n,m} (-1)^m$ to $Q_V$ type equation
\begin{eqnarray}
 && (u_{1,0} u_{0,1} u_{1,1} + u_{0,0} u_{0,1} u_{1,1} +
     u_{0,0} u_{1,0} u_{1,1} + u_{0,0} u_{1,0} u_{0,1}) k_2 + \nonumber	\\
 && (u_{0,0} + u_{1,0} + u_{0,1} + u_{1,1}) k_{12} = 0 . \label{u7}	
\end{eqnarray}

Consider  (\ref{u6}). If $k_2=k_{12}=0$, equation is degenerate:
\[
(T_2-1) (k_8 u_{1,0} / u_{0,0} ) = 0 .
\]
If at least one of $k_2$, $k_{12}$ is not zero, we can make $k_2 \ne 0$, using transformation $u_{n,m} = 1 / \hat u_{n,m}$. Then, using $u_{n,m} = \hat u_{n,m} + k_8 / (2 k_2)$, we make $k_8=0$. Eq. (\ref{u6}) with $k_8=0$ is reduced by the lattice-dependent transformation $u_{n,m} = \hat u_{n,m} (-1)^{n+m}$ to (\ref{u7}).

So the  Klein type equations are effectively equivalent to  $Q_V$.

\section*{Acknowledgments.}
RIY has been partially supported by the Russian Foundation for Basic Research (grant
numbers 10-01-00088-a and 11-01-00732-a). RIY thanks the Department of Electronic Engineering of Roma Tre University for its hospitality. LD  has been partly supported by the Italian Ministry of Education and Research, PRIN
``Nonlinear waves: integrable finite dimensional reductions and discretizations" from 2007
to 2009 and PRIN ``Continuous and discrete nonlinear integrable evolutions: from water
waves to symplectic maps" from 2010. We thank Christian Scimiterna for some illuminating discussions.


\end{document}